\documentclass[
intlimits,
sumlimits,
namelimits,
10pt,
a4paper,
]{article}

\usepackage{arxiv}

\theoremseparator{:}%

\usepackage{xifthen}

\usepackage{cmap}
\usepackage{textcomp}
\usepackage{mathtext}

\usepackage[T2A]{fontenc}
\usepackage[utf8]{inputenc}%
\usepackage[russian, main=english]{babel}

\usepackage{microtype}      % microtypography
\UseMicrotypeSet[protrusion]{alltext}

\ifx\No\undefined
    \def\No{\textnumero}
\fi

\usepackage{etex}

\usepackage{bm}
\usepackage{epigraph}
\usepackage{layout}

\usepackage{lscape}
\usepackage{epic}

\usepackage{soulutf8}

\usepackage{texnames}
\usepackage{paralist}

\usepackage{rotating}

\usepackage{wrapfig}

\usepackage[shortcuts]{extdash}

\usepackage{setspace}
\usepackage[perpage,bottom,multiple,stable]{footmisc}

\RequirePackage{color}
\usepackage{hyperref}
\hypersetup{backref,
linktoc=all
}
\hypersetup{pdfborder=0 0 0}

\hypersetup{pdfencoding=auto}

\ifthenelse{\boolean{xetex}}{}{%
  \usepackage{breakurl}
}

\definecolor{darkred}{rgb}{.5,0,0}
\definecolor{darkgoldenrod}{rgb}{.7,.5,0}
\definecolor{darkorange}{rgb}{.7,.4,0}
\definecolor{darkteal}{rgb}{0,.2,.2}
\definecolor{darkmagenta}{rgb}{.5,0,.5}
\definecolor{darkyellow}{rgb}{.5,.5,0}
\definecolor{darkgreen}{rgb}{0,.5,0}
\definecolor{darkblue}{rgb}{0,0,.5}
\definecolor{darkcyan}{rgb}{0,.5,.5}

\hypersetup{baseurl=http://, unicode, colorlinks, linkcolor=darkred, filecolor=darkteal, citecolor=darkgoldenrod, urlcolor=darkmagenta}

\usepackage{amsmath}
\usepackage{amssymb}
\usepackage{amsfonts}       % blackboard math symbols
\usepackage{nicefrac}       % compact symbols for 1/2, etc.
\usepackage{amscd}
\usepackage{slashed} % для обозначений по Фейнману "/"

\usepackage{accents}

\usepackage{mathtools}
\mathtoolsset{
showonlyrefs=true,
}

\allowdisplaybreaks

\RequirePackage{color}

\usepackage{pgfplotstable}
\pgfplotsset{compat=1.8}
\usepackage{booktabs}       % professional-quality tables
\usepackage{array}
\usepackage{colortbl}
\usepackage{multirow}

\usepackage{graphicx}

   \usepackage{float}
\graphicspath{{image/}}

\usepackage[scanall]{psfrag}

\usepackage{tikz}
\usetikzlibrary{arrows,shapes,graphs}

\usepackage{listingsutf8}
\ifthenelse{\boolean{luatex}\OR\boolean{xetex}}{}{%
   \lstset{inputencoding=utf8/koi8-r}}

\usepackage{calc}

\usepackage{url}

\usepackage{cite}

\ifdefined\refname\ifx\refname\empty\renewcommand*{\refname}{Литература}\else\renewcommand{\refname}{Литература}\fi\else\newcommand*{\refname}{Литература}\fi
\ifdefined\abstractname{\renewcommand{\abstractname}{}}\fi

\ifdefined\sectionFontShape{\renewcommand{\sectionFontShape}{\bfseries}}\else{\newcommand{\sectionFontShape}{\bfseries}}\fi
\ifdefined\subsectionFontShape{\renewcommand{\subsectionFontShape}{\bfseries}}\else{\newcommand{\subsectionFontShape}{\bfseries}}\fi
\ifdefined\subsubsectionFontShape{\renewcommand{\subsubsectionFontShape}{\bfseries}}\else{\newcommand{\subsubsectionFontShape}{\bfseries}}\fi
\ifdefined\paragraphFontShape{\renewcommand{\paragraphFontShape}{\bfseries}}\else{\newcommand{\paragraphFontShape}{\bfseries}}\fi
\ifdefined\subparagraphFontShape{\renewcommand{\subparagraphFontShape}{\bfseries}}\else{\newcommand{\subparagraphFontShape}{\bfseries}}\fi
\ifdefined\fi
\begin{document}

  {\selectlanguage{english}

    %% Подключение файла с титульным листом
    % титульная часть

% \crop[cam,noinfo,axes]

% Универсальный десятичный классификатор
% \udc{537.8}

\title[The generalized quantum mechanics of photon]{%
  The generalized quantum mechanics of Einstein \guillemotleft deinterlaced\guillemotright\ photon\\
  and Casimir force
}

\author[a]{Beilinson A.~A.}

\address[a]{
  Department of Theoretical Physics\\
  Peoples' Friendship University of Russia\\
  Russia, 117198, Moscow, Miklukho-Maklaya st., 6
}

\email[work]{alalbeyl@gmail.com}

% \thanks{%
  %
% }

\begin{abstract}
  \indent Based on Majorana equations the e.\nobreak-m. field as initially quantum object having isomorphic representation as quantum field of \guillemotleft deinterlaced\guillemotright\ photon is considered.
  
  \indent The calculation of Casimir force magnitude interpreted as consequence of an energy measurement of the generalized quantum field of a \guillemotleft deinterlaced\guillemotright\ photon in the state corresponding to a \guillemotleft Feynman path\guillemotright\ element is given. A metallic mirrors here plays role of classic apparatus measuring energy of this field.
\end{abstract}


% Here you can change the date presented in the paper title
%\date{September 9, 1985}
% Or remove it
\date{}

% Uncomment to override  the `A preprint' in the header
% \renewcommand{\headeright}{}
% \renewcommand{\undertitle}{}
% \def\keywordname{{\bfseries {Keywords and phrases}}}%
% \renewcommand{\shorttitle}{The Einstein photon and Casimir force}

%%% Add PDF metadata to help others organize their library
%%% Once the PDF is generated, you can check the metadata with
%%% $ pdfinfo template.pdf
\hypersetup{%
pdftitle={The generalized quantum mechanics of Einstein \guillemotleft deinterlaced\guillemotright\ photon and Casimir force},
pdfsubject={math-ph, quant-pm},
pdfauthor={Beilinson A.~A.},
pdfkeywords={generalized quantum mechanics, generalized path integral, Casimir force, Feynman paths, quantum field of a photon},
}

\renewcommand{\headrulewidth}{0pt}
\fancyheadoffset{0pt}
\lhead{}
\rhead{}
\chead{}
\lfoot{}
\rfoot{}
\cfoot{\thepage}
\copyrightinfo{}{}

\maketitle

%%% Local Variables:
%%% mode: latex
%%% coding: utf-8-unix
%%% TeX-master: "./default"
%%% End:
 %

    %% Подключение файла с аннотацией
    %% аннотация

% {\small

% }

% \indent {\bfseries Ключевые слова:}

%%% Local Variables:
%%% mode: latex
%%% coding: utf-8-unix
%%% TeX-master: "../default"
%%% End:%

    %% Подключение файла с оглавлением
    %% оглавление

% \tableofcontents

%%% Local Variables:
%%% mode: latex
%%% coding: utf-8-unix
%%% TeX-master: "../default"
%%% End:%

    %% Подключение файла с введением
    %% введение
\section*{Introduction}
% % \addtocontents{toc}{section}{Введение}
\label{sec:introduction}

\indent As known~(see~\cite{icikson:books:01:vol:01}), the Casimir forces are called the forces of attraction that arise between
neutral metallic mirrors at a small distance among them. These forces have the quantum-electrodynamic origin; despite the
smallness of them, their dependence on distance between mirrors (inverse proportionality of \(4\)th degree) has been
experimentally determined. Here the e.\nobreak-m. field in the article is understood as a fundamentally quantum object that has
no classic limit.

\begin{Note*}
  Majorana equations of e.\nobreak-m. field~(see~\cite{axiezer:book:1981:01}) not contain Planck constant~\(\hbar\) as well as the usual Maxwell equations. However the photon spin is not~\(1\), but is~\(\hbar\), which is why there should be a multiplier~\(\hbar\) in the l.h.s. and r.h.s. in the Majorana equations; these equations turn out to be equivalent to the previous ones only when~\(\hbar \neq 0\)~(for~\(\hbar = 0\) the Maxwell equations just disappear).

  \indent Therefore the Maxwell equations describes a quantum object not dependencies on quantity~\(\hbar \neq 0\) and therefore that has no classic analogue, so that all laws of classic e.\nobreak-d. is interpreted as means of such quantum field. This fundamental property the quantumness of e.\nobreak-m. field existing in two isomorphic forms representing physically radically different quantum field~(see~\cite{bejlinson:article:2023:02}). One of them (the \guillemotleft deinterlaced\guillemotright\ photon field) is responsible for the origin of Casimir force.
\end{Note*}

\indent In the article, Casimir force is interpreted as result of a energy macroscopic measurement of quantum e.\nobreak-m.
field of the \guillemotleft deinterlaced\guillemotright\ photon corresponding to small element of \guillemotleft Feynman path\guillemotright\ of the photon, in field of which a mirrors are placed, that are such a classic apparatus.

\indent Since the considered problem is obviously one-dimensionally the generalized Green function of arbitrary e.\nobreak-m. field in its usual and isomorphic forms is constructed and studied.

\indent In the article uses terminology and notation of monographs~\cite{gelfand:books:02:iss:01, gelfand:books:02:iss:02,
gelfand:books:02:iss:04}; in calculations, we assume the light velocity~\(c = 1\).

%%% Local Variables:
%%% mode: latex
%%% coding: utf-8-unix
%%% TeX-master: "../default"
%%% End:
%

    %% Подключение файла с основным содержанием
    %% основное содержание
% Обобщённая функция Грина одномерного квантового э.-м. поля как функционал на финитных функциях
\section{Generalized Green function of one-dimensional quantum e.\nobreak-m. field\\as functional on bump functions}
\label{sec:generalized_green_function_of_one-dimensional_quantum_em_field_as_functional_on_bump_functions}

\indent Recall that Maxwell equations of e.\nobreak-m. field in Majorana variables~\cite{axiezer:book:1981:01} for photon~\(\xi_{t}(x) = E_{t}(x) + i H_{t}(x)\) and antiphoton~\(\eta_{t}(x) = E_{t}(x) - i H_{t}(x)\) are become equations
\begin{equation}
  \label{eq:maxwell_equations_in_majorana_variables}
  \begin{aligned}
    & i\frac{\partial}{\partial{t}} \xi_{t}(x) = (S,\hat{p}) \xi_{t}(x) , & -i\frac{\partial}{\partial{t}} \eta_{t}(x) = (S,\hat{p}) \eta_{t}(x) ,
  \end{aligned}
\end{equation}
where~\(\hat{p} = \frac{1}{i} \nabla\), and~\(S\) is photon spin operators (infinitesimal operators of rotation around coordinate axes in representation corresponding to weight~\(1\),~see~\cite{gelfand:book:1972:09}).
\begin{equation}
  \label{eq:photon_spin_operators}
  \begin{aligned}
    & S_{1} =
    \begin{pmatrix*}[r]
      0 ,& 0 ,& 0 \\
      0 ,& 0 ,& -i \\
      0 ,& i ,& 0
    \end{pmatrix*}
    ,
    & S_{2} =
    \begin{pmatrix*}[r]
      0 ,& 0 ,& i \\
      0 ,& 0 ,& 0 \\
      -i ,& 0 ,& 0
    \end{pmatrix*}
    ,
    && S_{3} =
    \begin{pmatrix*}[r]
      0 ,& -i ,& 0 \\
      i ,& 0 ,& 0 \\
      0 ,& 0 ,& 0
    \end{pmatrix*}
    .
  \end{aligned}
\end{equation}
\(\xi_{t}(x)\), \(\eta_{t}(x)\) are supposed independent,~see~\cite{axiezer:book:1981:01}, that is why the Green function of these equations is represented by direct product of~\(3\times 3\) matrices and their solution is column of~\(6\) elements. Therefore, for the momentum representation of these equations are as follows:
\begin{equation}
  \label{eq:impulse_maxwell_equations_in_majorana_variables}
  \begin{aligned}
    & i\frac{\partial}{\partial{t}} \tilde{\xi}_{t}(p) = (S,p) \tilde{\xi}_{t}(p) , & -i\frac{\partial}{\partial{t}} \tilde{\eta}_{t}(p) = (S,p) \tilde{\eta}_{t}(p) .
  \end{aligned}
\end{equation}
We will consider a quantum e.\nobreak-m. field corresponding to states of a photon on the~\(z\) axis, in which the coordinates~\(x\),~\(y\) of photon is any (and hence the momentum~\(p_{x} = p_{y} = 0\) corresponding to them).

\indent The problem on the quantum e.\nobreak-m. field of an antiphoton is solved similarly. Almost to the end of the article will be investigated only a photon.

\indent Such e.\nobreak-m. fields in momentum representation, it is easy to see, satisfies the Shr\"{o}dinger equations
\begin{equation}
  \label{eq:impulse_equations_for_onedimensional_photon_motion}
  \begin{aligned}
    & i\frac{\partial}{\partial{t}}
    \begin{pmatrix*}[r]
      \tilde{\xi}_{t}^{(1)}(p_{z}) \\
      \tilde{\xi}_{t}^{(2)}(p_{z})
    \end{pmatrix*}
    = S_{z} p_{z}
    \begin{pmatrix*}[r]
      \tilde{\xi}_{t}^{(1)}(p_{z}) \\
      \tilde{\xi}_{t}^{(2)}(p_{z})
    \end{pmatrix*}
    ,
    & \frac{\partial}{\partial{t}} \tilde{\xi}_{t}^{(3)}(p_{z}) = 0 ,
  \end{aligned}
\end{equation}
the last component is \guillemotleft frozen\guillemotright\ and does not participate in evolution.

\indent It is clear that a generalized Green function of this equation is
\begin{equation}
  \label{eq:impulse_onedimensional_quantum_em_field_green_function}
  \tilde{M}_{t}(p_{z}) = \exp\!\left( t
  \begin{pmatrix*}[r]
    0 ,& -p_{z} \\
    p_{z} ,& 0
  \end{pmatrix*}
  \right) .
\end{equation}

\indent Bearing in mind the constructing of the coordinate representation the generalized Green function of this field as functional on columns of bump functions~\(\varphi(z) \in K\) (as Fourier preimage of solution in the momentum representation as functional~\(\int \bar{\tilde{M}}_{t} \begin{pmatrix*}[r] \psi_{1} \\ \psi_{2} \end{pmatrix*} \mathop{}\!{d{p_{z}}}\) on columns of analytic functions~\(\psi(p) \in Z\)~(см.~\cite{gelfand:books:01:iss:01}). Here
\begin{equation}
  \label{eq:parseval_equality_for_onedimensional_photon_motion}
  \int \bar{M}_{t}
  \begin{pmatrix*}[r]
    \varphi_{1} \\
    \varphi_{2}
  \end{pmatrix*}
  \mathop{}\!{d{z}} = \frac{1}{2 \pi} \int \bar{\tilde{M}}_{t}
  \begin{pmatrix*}[r]
    \psi_{1} \\
    \psi_{2}
  \end{pmatrix*}
  \mathop{}\!{d{p_{z}}}
\end{equation}
Let us consider the momentum representation of generalized Green function  of considered e.\nobreak-m. field of a photon in details.
\indent Remark that the momentum representation of the solution contains in power coefficient a Hermit matrix, therefore which can be reduced to diagonal view by unitary transform~\(\tilde{Q}(p_{z})\). Therefore we have
\begin{equation}
  \label{eq:impulse_green_function_for_onedimensional_photon_motion:diagonalization}
  \tilde{M}_{t}(p_{z}) = \exp\!\left( -i
  \begin{pmatrix*}[r]
    0 ,& -itp_{z} \\
    itp_{z} ,& 0
  \end{pmatrix*}
  \right) = \tilde{Q}^{+}(p_{z})
  \begin{pmatrix*}[r]
    \exp(-it|p_{z}|) ,& 0 \\
    0 ,& \exp(it|p_{z}|)
  \end{pmatrix*}
  \tilde{Q}(p_{z}) ,
\end{equation}
where
\begin{equation}
  \label{eq:impulse_foldy_wouthuysen operator_for_onedimensional_photon_motion_green_function}
  \tilde{Q}(p_{z}) = \frac{1}{\sqrt{2}}
  \begin{pmatrix*}[r]
    -i \operatorname{sgn} p_{z} ,& 1 \\
    i \operatorname{sgn} p_{z} ,& 1
  \end{pmatrix*}
  .
\end{equation}
But the Fourier preimages of numeric functionals~\(\exp(it|p_{z}|)\) and~\(\operatorname{sgn} p_{z}\) on~\(Z\) are the quantum Cauchy functional~(see~\cite{bejlinson:article:inproc:2018:01, bejlinson:article:inproc:2018:02})
\begin{equation}
  \label{eq:onedimensional_quantum_cauchy_functional}
  C_{it}(z) = \frac{1}{2} \left( \delta(t - z) + \delta(t + z) \right) + \frac{i}{2 \pi} \left( \frac{1}{t - z} + \frac{1}{t + z} \right)
\end{equation}
and, accordingly, the functional~\(-\frac{i}{\pi z}\)~(see~\cite{gelfand:books:02:iss:01},~p.~360~formula~19) on bump functions~\(\varphi(z) \in K\).

\begin{Note*}
  Previously it was shown that quantum Cauchy process~(see~\cite{bejlinson:article:inproc:2018:02}) is the correct analytic continuation in time of the Cauchy process transition probability~\(\frac{1}{\pi} \cdot \frac{t}{t^{2} + z^{2}}\)~(see~\cite{gnedenko:inbook:1949:13}) from the real semiaxis to the imaginary axis. Remark that formula~\ref{eq:onedimensional_quantum_cauchy_functional} can also be obtained by a simple calculation using improper integrals.
\end{Note*}

\indent Hence, it easy to see, that
\begin{equation}
  \label{eq:impulse_green_function_for_onedimensional_photon_motion:other_view}
  \begin{gathered}
    \tilde{M}_{t}(p_{z}) = \frac{1}{2}
    \begin{pmatrix*}[r]
      \exp(-it|p_{z}|) + \exp(it|p_{z}|) ,& \operatorname{sgn}(p_{z}) \cdot 2i \sin(t|p_{z}|) \\
      -\operatorname{sgn}(p_{z}) \cdot 2i \sin(t|p_{z}|) ,& \exp(it|p_{z}|) + \exp(-it|p_{z}|)
    \end{pmatrix*}
    = \\
    =
    \begin{pmatrix*}[r]
      \cos(tp_{z}) ,& \sin(tp_{z}) \\
      -\sin(tp_{z}) ,& \cos(tp_{z})
    \end{pmatrix*}
    .
  \end{gathered}
\end{equation}

\indent Hence, it easy to see, that we have coordinate representation the generalized Green function~\(\int \bar{M}_{t} \begin{pmatrix*}[r] \varphi_{1} \\ \varphi_{2} \end{pmatrix*} \mathop{}\!{d{z}}\) of Maxwell--Majorana equations, where
\begin{equation}
  \label{eq:green_function_for_onedimensional_photon_motion:other_view}
  M_{t}(z) = \frac{1}{2}
  \begin{pmatrix*}[r]
    \delta(t - z) + \delta(t + z) ,& i\left( \delta(t - z) - \delta(t + z) \right) \\
    -i\left( \delta(t - z) - \delta(t + z) \right) ,& \delta(t - z) + \delta(t + z)
  \end{pmatrix*}
  .
\end{equation}

\indent Thus the generalized Green function of the studied e.\nobreak-m. field is concentrated in points~\(z = \pm{t}\), which are the wave front, describing the initial state evolution of the field concentrated in point~\(z = 0\).

\indent We shall return to relation~\eqref{eq:impulse_green_function_for_onedimensional_photon_motion:diagonalization} that has a physical interpretation.

\indent Indeed, the matrix diagonalization underlying this equality is interpreted as cut-off the spin interaction between components of the considered generalized e.\nobreak-m. field and the rise in result a new field with the momentum representation of the generalized Green function
\begin{equation}
  \label{eq:impulse_green_function_for_onedimensional_quantum_cauchy_field}
  \tilde{\mu}_{t}(p_{z}) =
  \begin{pmatrix*}[r]
    \exp(-it|p_{z}|) ,& 0 \\
    0 ,& \exp(it|p_{z}|)
  \end{pmatrix*}
\end{equation}
with \guillemotleft deinterlaced\guillemotright\ components. Here the unitary operator~\(\tilde{Q}(p_{z})\) realizes both \guillemotleft deinterlacing\guillemotright\ and \guillemotleft interlacing\guillemotright\ of the field components.

\indent Moreover the coordinate representation the generalized Green function of this field with \guillemotleft deinterlaced\guillemotright\ components is~\(\int \bar{\mu}_{t} \begin{pmatrix*}[r] \varphi_{1} \\ \varphi_{2} \end{pmatrix*} \mathop{}\!{d{z}}\)~(see~\eqref{eq:green_function_for_onedimensional_photon_motion:other_view}), where
\begin{equation}
  \label{eq:green_function_for_onedimensional_quantum_cauchy_field}
  \mu_{t}(z) =
  \begin{pmatrix*}[r]
    C_{-it}(z) ,& 0 \\
    0 ,& C_{it}(z)
  \end{pmatrix*}
  .
\end{equation}
Let us compare those generalized quantum e.\nobreak-m. fields with unitary equivalent Green functionals~\(M_{t}(z)\) and~\(\mu_{t}(z)\). Since
\begin{equation}
  \label{eq:green_functional_for_onedimensional_photon_motion:other_view}
  \int \bar{M}_{t}(z)
  \begin{pmatrix*}[r]
    \varphi_{1}(z) \\
    \varphi_{2}(z)
  \end{pmatrix*}
  \mathop{}\!{d{z}} = \frac{1}{2}
  \begin{pmatrix*}[r]
    \varphi_{1}(t) + \varphi_{1}(-t) + i\left( \varphi_{2}(t) - \varphi_{2}(-t) \right) \\
    -i\left( \varphi_{1}(t) - \varphi_{1}(-t) \right) + \varphi_{2}(t) + \varphi_{2}(-t)
  \end{pmatrix*}
  ,
\end{equation}
then this functional is concentrated in points~\(z = \pm t\) of a quantum e.\nobreak-m. field wave front. At the same time the functional
\begin{equation}
  \label{eq:green_functional_for_onedimensional_photon_motion:foldy_wouthuysen_representation}
  \int \bar{\mu}_{t}(z)
  \begin{pmatrix*}[r]
    \varphi_{1}(z) \\
    \varphi_{2}(z)
  \end{pmatrix*}
  \mathop{}\!{d{z}} = \frac{1}{2}
  \begin{pmatrix*}[r]
    \frac{\varphi_{1}(t) + \varphi_{1}(-t)}{2} + \frac{i}{2\pi} v.p. \int \left( \frac{\varphi_{1}(z)}{t - z} + \frac{\varphi_{1}(z)}{t + z} \right) \mathop{}\!{d{z}} \\
    \frac{\varphi_{2}(t) + \varphi_{2}(-t)}{2} - \frac{i}{2\pi} v.p. \int \left( \frac{\varphi_{2}(z)}{t - z} + \frac{\varphi_{2}(z)}{t + z} \right) \mathop{}\!{d{z}}
  \end{pmatrix*}
\end{equation}
is concentrated on the exterior of wave front (outside) in~\(R^{(1)}\), see~\ref{sec:physic_attribution_of_quantum_em_field_of_deinterlaced_photon_and_casimir_forces}rd section of present work.

\begin{Remark*}
  So far, a bump functions have played a role of coordinates, in which it turned out to be possible to work with Green functionals of generalized quantum e.\nobreak-m. fields. It turns out that with help of complex bump functions~\(\psi(z) = \varphi(z) + i\phi(z)\)~(\(\varphi(z), \phi(z) \in K\)) we can construct a Shr\"{o}dinger equations corresponding to Green functionals of quantum e.\nobreak-m. fields and therefore their physically interpreted solutions.
\end{Remark*}

\indent Indeed, the existence of integral
\begin{equation}
  \label{eq:onedimensional_em_field_sourcewise_function}
  \int \bar{M}_{t}(z - z_{0})
  \begin{pmatrix*}[r]
    \psi_{1}(z_{0}) \\
    \psi_{2}(z_{0})
  \end{pmatrix*}
  \mathop{}\!{d{z_{0}}} =
  \begin{pmatrix*}[r]
    \psi_{1}(z) \\
    \psi_{2}(z)
  \end{pmatrix*}
\end{equation}
interpreting as solution of a Cauchy problem for a Shr\"{o}dinger type equation with Hamiltonian
\begin{equation}
  \label{eq:onedimensional_em_field_hamiltonian}
  \hat{\mathrm{H}} =
  \begin{pmatrix*}[r]
    0 ,& i\frac{\partial}{\partial z} \\
    -i\frac{\partial}{\partial z} ,& 0
  \end{pmatrix*}
\end{equation}
(Cf.~\eqref{eq:maxwell_equations_in_majorana_variables}).

\indent From this point on to simplify the recording it will be considered one of two independent components of the quantum e.\nobreak-m. field with Shr\"{o}dinger equation and its physically interpreted solution

\indent About similar procedure for~\(\mu_{t}(z)\) and~\(C_{it}(z)\) see~\ref{sec:physic_attribution_of_quantum_em_field_of_deinterlaced_photon_and_casimir_forces}rd section of present work.

\begin{Note*}
  The passage from~\(M_{t}(z)\) to~\(\mu_{t}(z)\) by means of unitary (and therefore isomorphic) transform is the fundamental rebuilding of the quantum e.\nobreak-m. field, representing its a completely new face. If the solution representation in the wave form~\eqref{eq:green_function_for_onedimensional_photon_motion:other_view} was known, then the constructed isomorphic solution~\eqref{eq:green_function_for_onedimensional_quantum_cauchy_field} has a diffuse character that allowed to discover and construct the generalized quantum measure in the space of a photon \guillemotleft Feynman paths\guillemotright\ with Hilbert instantaneous velocities~(see~\cite{bejlinson:article:inproc:2018:02}).
\end{Note*}

% Обобщенный функциональный интеграл, отвечающий обобщённой функции Грина построенного квантового э.-м. поля расплетенного фотона
\section{Generalized functional integral\\corresponding to generalized Green function\\of constructed quantum e.\nobreak-m. field of \guillemotleft deinterlaced\guillemotright\ photon}
\label{sec:generalized_functional_integral_corresponding_to_generalized_green_function_of_constructed_quantum_em_field_of_deinterlaced_photon}

\indent Recall that the generalized quantum Cauchy process~\(C_{it}(z)\) ---~see~\eqref{eq:green_function_for_onedimensional_photon_motion:other_view},~(and therefore the~\(\mu_{t}(z)\)) is continuable to a generalized countably additive complex measure in the space dual the space of Hilbert instantaneous velocities of a photon, that is turned out to be a compact part of the continuous function space,~see~\cite{bejlinson:article:inproc:2018:02, gelfand:books:01:iss:04}.

\indent The countably additivity of the quantum generalized Cauchy measure yields to possibility of quantum-theoretic expansion the state term of a photon to states on trajectories~(\guillemotleft Feynman paths\guillemotright), continuous trajectories with Hilbert derivative.

\indent We will assume that~\(\Delta t_{1}, \ldots, \Delta t_{n}\)~(\(t_{0} = 0\), \(t_{n} = t\)) is certain partitioning of a time interval~\([0, t]\), and~\(\Delta z_{j}\)~(\(j = 1, \ldots, n\)) is shifts of the photon at the appropriate time intervals.

\indent Using the kernel theorem~(see~\cite{gelfand:books:02:iss:04}) we have
\begin{equation}
  \label{eq:onedimensional_quantum_cauchy_functionals:direct_product:by_kernel_theorem}
  \int \left( \prod\nolimits_{j = 1}^{n} \bar{C}_{i \Delta t_{j}}(\Delta z_{j}) \right) \varphi(\Delta z_{1}, \ldots, \Delta z_{n}) \mathop{}\!{d{z_{j}} \ldots d{z_{n}}} ,
\end{equation}
where~\(\varphi(z_{1}, \ldots, z_{n})\) is bump functions of~\(n\) variables.

\indent This allows to interpret that functional as a generalized state of the photon located sequentially on~\(n\) segments~\(\Delta z_{j}\) at the appropriate time intervals~\(\Delta t_{j}\). This implies that the convolution of these states
\begin{equation}
  \label{eq:onedimensional_quantum_cauchy_functionals:convolution}
  \int \left( \prod\nolimits_{j = 1}^{n} \bar{C}_{i \Delta t_{j}}(\Delta z_{j}) \right) \varphi(\Delta z_{1} + \ldots + \Delta z_{n}) \mathop{}\!{d{z_{j}} \ldots d{z_{n}}} = \int \bar{C}_{it}(z) \varphi(z) \mathop{}\!{d{z}}
\end{equation}
is the state of the photon at the last moment of time.

\indent The existence of the generalized Cauchy measure gives possibility of passage to the limit in the writing of the quantum Cauchy process and passage to the generalized functional functional integral on \guillemotleft Feynman paths\guillemotright
\begin{multline}
  \label{eq:onedimensional_quantum_cauchy_functional_path_integral:passage_to_the_limit}
  \lim_{\max \Delta t_{j} \to 0} \int \left( \prod\nolimits_{j = 1}^{n} \bar{C}_{i \Delta t_{j}}(\Delta z_{j}) \right) \varphi(\Delta z_{1} + \ldots + \Delta z_{n}) \mathop{}\!{d{z_{j}} \ldots d{z_{n}}} =\\
  = \int_{\{z_{\tau}\}} \left( \prod\nolimits_{\tau = 0}^{t} \bar{C}_{i d{\tau}}( d{z(\tau)}) \right) \varphi[z(\tau)] \mathop{}\!{\prod\nolimits_{\tau = 0}^{t} d{z_{\tau}}} = \int \bar{C}_{it}(z) \varphi(z) \mathop{}\!{d{z}} ,
\end{multline}
where~\(\{z_{\tau}\}\) is the support of generalized quantum Cauchy measure ---~\guillemotleft Feynman paths\guillemotright\ ---~set of continuous trajectories on~\([0, t]\), which are compact in topology of uniform convergence,~\(d{z_{\tau}}\) is differential at constant time,~\(d{z(\tau)} = \dot{z}(\tau) d{\tau}\)~(\(\dot{z}(\tau) \in L_{2}(0, t)\)), and~\(\varphi[z(\tau)]\) are bump functionals on~\(\{z_{\tau}\}\).

\indent Remarkably, those solution forms of the quantum fields, belonging to isomorphic Hilbert spaces with common scalar product, that allowed to discover the existence of such measure with similar properties in usual form of the quantum e.\nobreak-m. field, on average (in eikonal approximation) giving the equations of classic electrodynamics.

% Энергия квантового э.-м. поля и силы Казимира
\section{Physic attribution of quantum e.\nobreak-m. field of \guillemotleft deinterlaced\guillemotright\ photon\\and Casimir forces}
\label{sec:physic_attribution_of_quantum_em_field_of_deinterlaced_photon_and_casimir_forces}

\indent We consider the structure of the quantum field of \guillemotleft deinterlaced\guillemotright\ photon with the generalized Green function~\(C_{it}(z)\) (see~\eqref{eq:onedimensional_quantum_cauchy_functional}) with the Hamilton functional~\(\int \hat{\bar{\mathrm{H}}}(z) \psi_{0}(z) \mathop{}\!{d{z}} = \frac{1}{\pi} \int z^{-2} \psi_{0}(z) \mathop{}\!{d{z}}\) and integral Shr\"{o}dinger equation
\begin{equation}
  \label{eq:onedimensional_shrodinger_equation:deinterlaced_photon_field}
  i\frac{\partial}{\partial t}\psi_{t}(z) = \frac{1}{\pi} \int \alpha^{-2} \psi_{0}(z - \alpha) \mathop{}\!{d{\alpha}} .
\end{equation}
with Cauchy problem solution on each small time interval~\(\Delta t\)
\begin{equation}
  \label{eq:onedimensional_deinterlaced_photon_field_structure_on_small_time_interval}
  \psi_{\Delta t}(z) = \psi_{0}(z) + i\Delta t \int \hat{\bar{\mathrm{H}}}(z - \alpha) \psi_{0}(\alpha) \mathop{}\!{d{\alpha}} - i\Delta t \int \hat{\bar{\mathrm{H}}}(\alpha) \psi_{0}(z - \alpha) \mathop{}\!{d{\alpha}}
\end{equation}
Taking into account the definition of functional~\(\alpha^{-2}\)
\begin{equation}
  \label{eq:principal_value_integral:deinterlaced_photon_field}
  v.p. \int \alpha^{-2} \psi_{0}(z - \alpha) \mathop{}\!{d{\alpha}} = \lim_{\varepsilon \to 0} \left( \int\nolimits_{-\infty}^{-\varepsilon} + \int\nolimits_{\varepsilon}^{\infty} \right) \alpha^{-2} \psi_{0}(z - \alpha) \mathop{}\!{d{\alpha}}
\end{equation}
where~\(\varepsilon\) is any (see~\cite{gelfand:books:02:iss:01},~p.~52~formula~7), in our problem, where~\(\varepsilon > 0\), we have
\begin{equation}
  \label{eq:principal_value_integral:deinterlaced_photon_field:our_case}
  v.p. \int \alpha^{-2} \psi_{0}(z - \alpha) \mathop{}\!{d{\alpha}} = \left. \left( \int\nolimits_{-\infty}^{-\varepsilon} + \int\nolimits_{\varepsilon}^{\infty} \right) \alpha^{-2} \psi_{0}(z - \alpha) \mathop{}\!{d{\alpha}} \right|_{\varepsilon \to 0} ,
\end{equation}
where different from zero the result is obtained only on the even by~\(\alpha\) the bump functions~\(\psi_{0}(z - \alpha)\).

\indent Therefore, we have the solution of Cauchy problem for Shr\"{o}dinger equation on small time interval~\(\Delta t\)
\begin{equation}
  \label{eq:onedimensional_deinterlaced_photon_field_shrodinger_equation_solution_on_small_time_interval}
  \psi_{\Delta t}(z) = \psi_{0}(z) - \left. \frac{i\Delta t}{\pi} \left( \int\nolimits_{-\infty}^{-\varepsilon} + \int\nolimits_{\varepsilon}^{\infty} \right) \alpha^{-2} \psi_{0}(z - \alpha) \mathop{}\!{d{\alpha}} \right|_{\varepsilon \to 0}
\end{equation}
Note, that \guillemotleft deinterlaced\guillemotright\ photon, according to this formula, being in any localized state~\(\psi_{0}(z)\) at~\(t = 0\), at the very first moment it goes beyond the limits of the classic light cone~\(z = t\),~\(t < \Delta t\), on which it was located at~\(t = 0\), and filling at once whole coordinate space outside the small \(\varepsilon\)\nobreak-vicinity of the origin of coordinates.

\indent Therefore the Hamiltonian of such functional is generalized function
\begin{equation}
  \label{eq:onedimensional_em_photon_field_hamiltonian:deinterlaced:epsilon_vicinity}
  \hat{\mathrm{H}}_{\varepsilon}(z) = \frac{1}{\pi}
  \begin{cases}
    z^{-2} ,& -\infty < z < -\varepsilon \\
    \shoveleft{0 ,}& -\varepsilon \leqslant z \leqslant \varepsilon \\
    z^{-2} ,& \varepsilon < z < \infty
  \end{cases}
\end{equation}
Remark, that the constructed solution~\(\psi_{\Delta t}(z)\) conveniently to  take as the \guillemotleft deinterlaced\guillemotright\ photon state, in which mirrors are introduced, that are mentioned in \guillemotleft Introduction\guillemotright, at points~\(\pm\varepsilon\), since such mirrors does not deform such quantum field.

\indent Thus the wave functional, in which mirrors are introduced,
\begin{equation}
  \label{eq:onedimensional_em_photon_field_functional:deinterlaced:with_mirrors}
  \Delta_{t} \psi_{0}(z) = \left. \frac{-i\Delta t}{\pi} \left( \int\nolimits_{-\infty}^{-\varepsilon} + \int\nolimits_{\varepsilon}^{\infty} \right) \alpha^{-2} \psi_{0}(z - \alpha) \mathop{}\!{d{\alpha}} \right|_{\varepsilon \to 0} = \left. \psi_{\varepsilon}(z) \right|_{\varepsilon \neq 0}
\end{equation}
and~\(2\varepsilon\) is distance between mirrors.

\indent Moreover, since the operator~\(\frac{\partial}{\partial\varepsilon}\) (meaning the simultaneous moving asunder the mirrors), acting on wave functional~\(\psi_{\varepsilon}(z)\)
\begin{equation}
  \label{eq:onedimensional_photon_functional:deinterlaced:differential_operator_by_epsilon_acting}
  \frac{\partial}{\partial\varepsilon} \frac{-i\Delta t}{\pi} \left( \int\nolimits_{-\infty}^{-\varepsilon} + \int\nolimits_{\varepsilon}^{\infty} \right) \alpha^{-2} \psi_{0}(z - \alpha) \mathop{}\!{d{\alpha}} = \frac{i\Delta t}{\pi} (\accentset{\leftarrow}{\varepsilon}{\,}^{-2} + \accentset{\rightarrow}{\varepsilon}{\,}^{-2})\alpha^{-2} \psi_{0}(z - \alpha)
\end{equation}
(arrows mean direction of forces acting on left and right mirror) has the meaning of the force operator acting on mirrors from the side of the quantum field (Casimir forces) and moving asunder the mirrors. Here the a negative energy of the quantum field of \guillemotleft deinterlaced\guillemotright\ photon corresponds to Casimir forces, contrary to positive energy of the quantum field of usual photon.

\indent Remark, that the operator~\(\frac{\partial}{\partial\varepsilon}\) acts on an element of the Hilbert space~\(\psi_{\varepsilon}(z)\)~(see~\eqref{eq:onedimensional_em_photon_field_functional:deinterlaced:with_mirrors}), is interpreted as a projection operator of this state to one of summand (mutually non-orthogonal vectors of Hilbert space) of integral sum.
                  
\indent Therefore the average of Casimir operator~\(\frac{\partial}{\partial\varepsilon}\) on quantum field in the state~\(\psi_{\varepsilon}(z)\) is
\begin{equation}
  \label{eq:casimir_operator_mean:deinterlaced:on_epsilon_state}
  \overline{\frac{\partial}{\partial\varepsilon}} = \frac{\int \bar{\psi_{\varepsilon}}(z) \frac{\partial}{\partial\varepsilon} \psi_{\varepsilon}(z) \mathop{}\!{d{z}}}{\int \bar{\psi_{\varepsilon}}(z) \psi_{\varepsilon}(z) \mathop{}\!{d{z}}} = \frac{\int \bar{\psi_{\varepsilon}}(z) (\accentset{\leftarrow}{\varepsilon}{\,}^{-2} + \accentset{\rightarrow}{\varepsilon}{\,}^{-2}) \psi_{0}(z - \varepsilon) \mathop{}\!{d{z}}}{\int \bar{\psi_{\varepsilon}}(z) \psi_{\varepsilon}(z) \mathop{}\!{d{z}}} .
\end{equation}
For calculation of this fraction we will write the integrand vectors in the trigonometric (orthonormal) basis. Since
\begin{equation}
  \label{eq:hilbert_vectors:deinterlaced:orthonormal_basis}
  \int \bar{\psi_{0}}(z - \alpha) \psi_{0}(z - \alpha^{'}) \mathop{}\!{d{z}} = \int \exp\!\left( -ip(\alpha - \alpha^{'}) \right) \left| \tilde{\psi}_{0}(p) \right|^{2} \mathop{}\!{d{p}} , 
\end{equation}
we have~(see~\eqref{eq:onedimensional_photon_functional:deinterlaced:differential_operator_by_epsilon_acting})
\begin{equation}
  \label{eq:casimir_operator_mean:deinterlaced:on_epsilon_state:calculation}
  \begin{gathered}
    \overline{\frac{\partial}{\partial\varepsilon}} = (\accentset{\leftarrow}{\varepsilon}{\,}^{-2} + \accentset{\rightarrow}{\varepsilon}{\,}^{-2}) \frac{\int \left( \int\nolimits_{-\infty}^{-\varepsilon} + \int\nolimits_{\varepsilon}^{\infty} \right) \alpha^{-2} \exp\!\left( -ip(\alpha - \varepsilon) \right) \left| \tilde{\psi}_{0}(p) \right|^{2} \mathop{}\!{d{\alpha}} \mathop{}\!{d{p}}}{\int \left( \int\nolimits_{-\infty}^{-\varepsilon} + \int\nolimits_{\varepsilon}^{\infty} \right) \alpha^{-2} \exp(-ip\alpha) \left| \tilde{\psi}_{0}(p) \right|^{2} \left( \int\nolimits_{-\infty}^{-\varepsilon} + \int\nolimits_{\varepsilon}^{\infty} \right) {\alpha^{'}}^{-2} \exp(-ip\alpha^{'}) \mathop{}\!{d{\alpha}} \mathop{}\!{d{\alpha^{'}}} \mathop{}\!{d{p}}} =\\
    = (\accentset{\leftarrow}{\varepsilon}{\,}^{-2} + \accentset{\rightarrow}{\varepsilon}{\,}^{-2}) \frac{\int \left( \int\nolimits_{-\infty}^{-\varepsilon} + \int\nolimits_{\varepsilon}^{\infty} \right) \alpha^{-2} \exp\!\left( -ip(\alpha - \varepsilon) \right) \left| \tilde{\psi}_{0}(p) \right|^{2} \mathop{}\!{d{\alpha}} \mathop{}\!{d{p}}}{\int \left| \left( \int\nolimits_{-\infty}^{-\varepsilon} + \int\nolimits_{\varepsilon}^{\infty} \right) \alpha^{-2} \exp(-ip\alpha) \mathop{}\!{d{\alpha}} \right|^{2} \left| \tilde{\psi}_{0}(p) \right|^{2} \mathop{}\!{d{p}}} =\\
    = (\accentset{\leftarrow}{\varepsilon}{\,}^{-2} + \accentset{\rightarrow}{\varepsilon}{\,}^{-2}) \frac{\int\nolimits_{0}^{\infty} \int\nolimits_{\varepsilon}^{\infty} \alpha^{-2} \cos(p\alpha) \cos(p\varepsilon) \left| \tilde{\psi}_{0}(p) \right|^{2} \mathop{}\!{d{\alpha}} \mathop{}\!{d{p}}}{2\int\nolimits_{0}^{\infty}  \left( \int\nolimits_{\varepsilon}^{\infty} \alpha^{-2} \cos(p\alpha) \mathop{}\!{d{\alpha}} \right)^{2} \left| \tilde{\psi}_{0}(p) \right|^{2} \mathop{}\!{d{p}}}
  \end{gathered}
\end{equation}
Since this fraction is interpreted in the geometry of a Hilbert space as the square cosine of the angle between the vectors of this space~\(\frac{\partial}{\partial\varepsilon} \psi_{\varepsilon}(z)\) and~\(\psi_{\varepsilon}(z)\), the value of this fraction does not depend on the choice of bump function~\(\psi_{0}(z)\), that is why any bump function can be used to numerically find the value of this fraction, for example,~\(\psi_{0}(z) = \exp(-z^{2})\).

\indent Recall that the real quantum e.\nobreak-m. field in the state of \guillemotleft deinterlaced\guillemotright\ photons (taking into account the states of photon-antiphoton) contains two independent components.

\indent Recall also that the influence of the field corresponding to element~\(dt\) of a \guillemotleft Feynman path\guillemotright~\(z(t)\) is taken into account, while the whole plane~(\(z = 0\)) is exist, which is why we have that \guillemotleft Casimir forces\guillemotright corresponding each area unit of the plane~\(z = 0\) is
\begin{equation}
  \label{eq:casimir_operator_mean:deinterlaced:on_epsilon_state:area_unit}
  \left( 2\overline{\frac{\partial}{\partial\varepsilon}} \right)^{-2} .
\end{equation}
 
\indent At the same time, there is a continuous \guillemotleft tissue\guillemotright\ of the quantum field~\(\psi_{0}(z)\) of \guillemotleft deinterlaced\guillemotright\ photon, where the Casimir forces do not appears absolutely. That is interpreted as the presence of negative pressure forces into this environment stretching that \guillemotleft tissue\guillemotright. Therefore the Casimir forces measured in known experiment are exactly equal and opposite to the considered ones in present work.

%%% Local Variables:
%%% mode: latex
%%% coding: utf-8-unix
%%% TeX-master: "./default"
%%% End:
%

    %% Подключение файла с заключением
    %% заключение
\section*{Conclusion}
\label{sec:conclusion}

\indent The work shows that the real existence of Casimir forces indicates the reality of the existence of two unitary equivalent, but fundamentally different physically forms of the quantum e.\nobreak-m. field, one of which (the known) is described by the Maxwell--Majorana equations with the solutions inside the light cone, contrary to other form --- the quantum field, the field of \guillemotleft deinterlaced\guillemotright\ photon, existing outside the light cone and responsible for Casimir forces.

%%% Local Variables:
%%% mode: latex
%%% coding: utf-8-unix
%%% TeX-master: "../default"
%%% End:
%

    %% Подключение файла со списком публикаций
    %% список литературы

% \section*{\refname}
% \addtocontents{toc}{section}{Список литературы}
\label{sec:references}

% \bibliographystyle{ugost2008ls}
% \putbib[bib/alt]
\bibliography{default}

\begin{thebibliography}{10}
\def\selectlanguageifdefined#1{
\expandafter\ifx\csname date#1\endcsname\relax
\else\selectlanguage{#1}\fi}
\providecommand*{\href}[2]{{\small #2}}
\providecommand*{\url}[1]{{\small #1}}
\providecommand*{\BibUrl}[1]{\url{#1}}
\providecommand{\BibAnnote}[1]{}
\providecommand*{\BibEmph}[1]{#1}
\ProvideTextCommandDefault{\cyrdash}{\iflanguage{russian}{\hbox
  to.8em{--\hss--}}{\textemdash}}
\providecommand*{\BibDash}{\ifdim\lastskip>0pt\unskip\nobreak\hskip.2em plus
  0.1em\fi
\cyrdash\hskip.2em plus 0.1em\ignorespaces}
\renewcommand{\newblock}{\ignorespaces}

\bibitem{axiezer:book:1981:01}
\selectlanguageifdefined{english}
\BibEmph{Axiezer~A.~I., Beresteckij~V.~B.} Quantum electrodynamics~/ Ed.\ by\
  L.~P.~Rusakova. \BibDash
\newblock 4th edition. \BibDash
\newblock M.~: Nauka, 1981.

\bibitem{bejlinson:article:inproc:2018:02}
\selectlanguageifdefined{english}
\BibEmph{Beilinson~A.~A.} The Cauchy process in imaginary time and the path
  integrals in relativistic quantum mechanics of Dirac's electrons and
  Einstein's photons~// International Scientific Journal. \BibDash
\newblock Vol.~4 of \BibEmph{Journal of Mathematics}. \BibDash
\newblock AU~: CreateSpace, 2018. \BibDash Aug. \BibDash
\newblock P.~103--116.

\bibitem{bejlinson:article:inproc:2018:01}
\selectlanguageifdefined{english}
\BibEmph{Beilinson~A.~A.} The quantum mechanics of the free Dirac electrons and
  Einstein photons, and the Cauchy distribution~// International Scientific
  Journal. \BibDash
\newblock Vol.~4 of \BibEmph{Journal of Mathematics}. \BibDash
\newblock AU~: CreateSpace, 2018. \BibDash Aug. \BibDash
\newblock P.~36--55.

\bibitem{bejlinson:article:2023:02}
\selectlanguageifdefined{english}
\BibEmph{Beilinson~A.~A.} On certain paradoxical properties of Einstein photon
  fields and Dirac electron-positron fields. \BibDash
\newblock 2023. \BibDash
\newblock 2305.03717.

\bibitem{gelfand:book:1972:09}
\selectlanguageifdefined{english}
\BibEmph{Gel'fand~I.~M., Minlos~R.~A., Shapiro~Z.~Ya.} The representations of
  rotation group and of Lorentz group, their application. \BibDash
\newblock M.~: Fizmatgiz, 1972.

\bibitem{gelfand:books:01:iss:01}
\selectlanguageifdefined{english}
\BibEmph{Gel'fand~I.~M., Shilov~G.~E.} Generalized functions and operations on
  them. Generalized functions no. 1st. \BibDash
\newblock 2nd edition. \BibDash
\newblock M.~: Fizmatgiz, 1959.

\bibitem{gelfand:books:02:iss:01}
\selectlanguageifdefined{english}
\BibEmph{Gel'fand~I.~M., Shilov~G.~E.} Generalized functions: Properties and
  Operations~/ Ed.\ by\ ~trans. from~rus.~E.~Saletan. \BibDash
\newblock N.~Y. and L.~: Academic Press, 1964. \BibDash
\newblock Vol.~1st of \BibEmph{Generalized functions}.

\bibitem{gelfand:books:02:iss:02}
\selectlanguageifdefined{english}
\BibEmph{Gel'fand~I.~M., Shilov~G.~E.} Spaces of Fundamental and Generalized
  Functions~/ Ed.\ by\ C.~P.~Peltzer~trans. from~rus.~M.~D.~Friedman,
  A.~Feinstein. \BibDash
\newblock N.~Y. and L.~: Academic Press, 1968. \BibDash
\newblock Vol.~2nd of \BibEmph{Generalized functions}.

\bibitem{gelfand:books:01:iss:04}
\selectlanguageifdefined{english}
\BibEmph{Gel'fand~I.~M., Vilenkin~N.~Ya.} Some questions of the application of
  harmonic analysis {$\bullet$} Rigged Hilbert spaces. Generalized functions
  no. 4th. \BibDash
\newblock M.~: Fizmatgiz, 1961.

\bibitem{gelfand:books:02:iss:04}
\selectlanguageifdefined{english}
\BibEmph{Gel'fand~I.~M., Vilenkin~N.~Ya.} Application of Harmonic Analysis~/
  Ed.\ by\ ~trans. from~rus.~A.~Feinstein. \BibDash
\newblock N.~Y. and L.~: Academic Press, 1964. \BibDash
\newblock Vol.~4th of \BibEmph{Generalized functions}.

\bibitem{gnedenko:inbook:1949:13}
\selectlanguageifdefined{english}
\BibEmph{Gnedenko~B.~V., Kolmogorov~A.~N.} The limit distributions for sums of
  independent random variables. \BibDash
\newblock M.~: GIZT-TL, 1949.

\bibitem{icikson:books:01:vol:01}
\selectlanguageifdefined{english}
\BibEmph{Itzykson~C., Zuber~J.~B.} Quantum field theory~/ Ed.\ by\ ~trans.
  from~eng.~R.~M.~Mir~Kasimov. \BibDash
\newblock M.~: «Mir», 1984. \BibDash
\newblock Vol.~1.

\end{thebibliography}
%\input default.bbl

%%% Local Variables:
%%% mode: latex
%%% coding: utf-8-unix
%%% TeX-master: "../default"
%%% End:%

    %% Подключение файла с приложением
    % приложение
% \section{Приложение}
% \label{sec:application}

%%% Local Variables:
%%% mode: latex
%%% coding: utf-8-unix
%%% TeX-master: "./default"
%%% End:%

  } % END \selectlanguage

\end{document}